\documentclass[showpacs,twocolumn,preprintnumbers,amsmath,amssymb,floatfix]{revtex4}
\usepackage{epsfig}
\usepackage{graphics}
\usepackage{color}

\begin{document}

\title{Trapping of Neutral Rubidium with a Macroscopic Three-Phase Electric Trap}
\author{T. Rieger}
\affiliation{Max-Planck-Institut f\"ur Quantenoptik,
Hans-Kopfermann-Str. 1, D-85748 Garching, Germany}
\author{P. Windpassinger}
\altaffiliation{Present address: QUANTOP, Niels Bohr Institute,
DK-2100 Copenhagen, Denmark}
\author{S.A. Rangwala}
\altaffiliation{Present address: Raman Research Institute, C. V.
Raman Avenue, Sadashivanagar, Bangalore 560080, India}
\author{G. Rempe}
\author{P.W.H. Pinkse}
\affiliation{Max-Planck-Institut f\"ur Quantenoptik,
Hans-Kopfermann-Str. 1, D-85748 Garching, Germany}
\date{\today, PREPRINT}

\begin{abstract}

We trap neutral ground-state rubidium atoms in a macroscopic trap
based on purely electric fields. For this, three electrostatic field
configurations are alternated in a periodic manner. The rubidium is
precooled in a magneto-optical trap, transferred into a magnetic
trap and then translated into the electric trap. The electric trap
consists of six rod-shaped electrodes in cubic arrangement, giving
ample optical access. Up to $10^5$ atoms have been trapped with an
initial temperature of around 20 microkelvin in the three-phase
electric trap. The observations are in good agreement with detailed
numerical simulations.


\end{abstract}

\maketitle

\vspace{1cm}

Trapping is essential in many modern experiments in physics. A trap
allows the interaction time of the trapped species with other
particles or fields to be greatly extended. Trapping enabled
breakthrough experiments with ions, atoms and molecules. Each newly
demonstrated trap paved the way for new classes of experiments. Thus
far, electric traps for neutral polarizable particles have received
relatively little attention. They have first been proposed for
excited (Rydberg) atoms~\cite{WingPRL80} and, later, for
ground-state atoms~\cite{Shimizu92,Morinaga94}. Ground-state
particles lower their energy in electric fields. Hence these
particles are attracted by high fields (``high-field seekers'').
Since Maxwell's equations don't allow the creation of an
electrostatic maximum in free space, time-dependent
(pseudo-electrostatic or ``AC'') fields are required for trapping, a
principle well known from ion traps. Only recently,
two-dimensional~\cite{Junglen04} and
three-dimensional~\cite{Veldhoven05} versions of AC traps have been
demonstrated with cold polar molecules having a large and linear
Stark shift. For polar molecules, these traps offer the advantage of
a deep trapping potential, which can trap both low- and high-field
seeking states. Using laser-cooled strontium atoms, the Katori group
has demonstrated a chip version of an AC electric trap with the
motivation of performing precision spectroscopy for
metrology~\cite{Kishimoto06}.

In this Letter we report on an experiment where laser-cooled
rubidium atoms are trapped in a macroscopic AC electric trap. This
result opens the perspective of confining cold molecules and atoms
in the same spatial region for the purpose of using the optically
cooled atoms as a coolant for the molecules. For this, a large
volume of several mm$^3$, ample optical access and a relatively deep
effective potential depth are essential. The geometry of our trap is
in essence the one proposed in \cite{Shimizu92,Morinaga94} with
rectangular driving. Our trap is a three-phase trap, i.e., a full
cycle of its operation can be divided in three different phases. The
three-phase trap offers cubic symmetry rather than the cylindrical
one of three-dimensional two-phase traps~\cite{Peik99}. Therefore,
the average restoring force is isotropic about the trap
center~\cite{Wuerker59}.

\begin{figure}[htb]\begin{center}
\epsfig{file=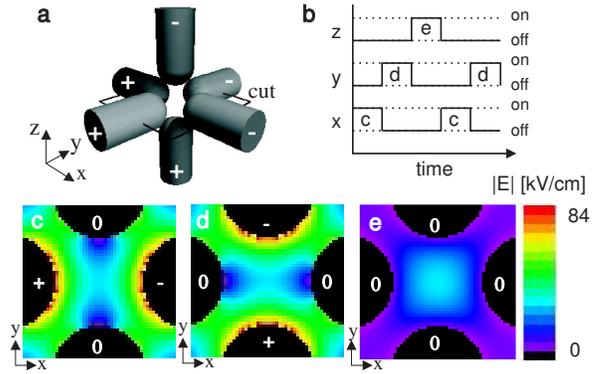,width=0.43\textwidth}%
\caption{(color online) The AC electric trap. a) the 6 electrodes.
b) the switching diagram, indicating that in each full cycle each
pair of electrodes is switched on for one third of the time. c)-e)
the absolute value of the electric field in the $x-y$ plane in the
three consecutive phases of a full driving cycle with electrodes at
$\pm7\,$kV.}
\label{ACidea}%
\end{center}\end{figure}

Our trap is depicted in Fig.~\ref{ACidea}. It consists of three
pairs of electrodes in a cubic arrangement. Trapping is achieved by
alternating the electric field configuration in a cyclic manner with
driving frequency $f$. The phase of the driving field is $\phi=2\pi
t/T \mod 2\pi$, with $t$ the time and $T=1/f$ the period. At each
point of time, only two opposing electrodes are given large voltages
of opposite sign; the other four electrodes are grounded. This leads
to an electric field of which the absolute value $|E|$ has the shape
of a saddle, increasing towards the two charged electrodes and
decreasing in the two perpendicular directions. In the following we
first focus on the field near the trap center. Assuming that in the
first phase of the driving cycle the electrodes in the x direction
are charged, the electric field strength can here be approximated in
second order in the spatial coordinates by
\begin{equation}
    |E|=E_0 + b x^2-\frac{1}{2}b(y^2+z^2) \ \ 0\le\phi<2\pi/3,
    \label{HarmonicPot}
\end{equation}
where in our experiment $E_0=30.4\,$kV/cm and $b=19.8\ $
kV/cm/mm$^2$ are coefficients found from the full numerical field
calculation~\cite{Simion} with the $x$ electrodes at $\pm7\,$kV.
Atoms experience a quadratic Stark shift in this field: $W_{\rm
S}=-\frac{1}{2}\alpha |E|^2$, where $\alpha$ is the atom's static
polarizability~\cite{RbPolarizability}. Due to the gradients in
$|E|$, the field exerts a force ${\mathbf F}=-\nabla W_{\rm S}$ on
the atoms. This force pushes the atoms towards the high electric
fields near the surfaces of the charged electrodes. This defocusses
the atom cloud in the $x$ direction during $0\le\phi<2\pi/3$. At the
same time, the atoms are focussed in the $y$ and $z$ direction. In
the second phase of the trap cycle ($2\pi/3\le\phi<4\pi/3$), the
pair of electrodes in the $y$ direction is charged while
simultaneously the $x$ and $z$ pair are grounded. Now the $y$
direction is defocussing, whereas the $x$ and $z$ direction are
focussing. In the third and last phase ($4\pi/3\le\phi<6\pi/3$), the
focussing and defocussing directions are permutated once more. With
the additional approximation that ${\mathbf F}=\alpha|E|\nabla
|E|\approx \alpha E_0\nabla |E|$, the force on an atom at a given
point in space averages to zero in a full cycle. However, under the
influence of the periodic driving the atom moves and the average
force over a full cycle is finite and directed to the trap center.
Assuming the drive frequency is high enough, the motion can be be
divided in two parts: the cyclic motion locked to the driving field
is called the micromotion; the motion under influence of the average
force is called the secular motion. For large drive frequencies one
can average over the micromotion to find the average restoring force
yielding the secular motion. For a driving frequency below a certain
cut-off value, the micromotion becomes unstable, the atoms hit the
electrodes and are lost.

The equations of motion derived from ${\mathbf F}\approx \alpha
E_0\nabla |E|$ with $|E|$ from Eq.\,(\ref{HarmonicPot}) can be
solved analytically~\cite{Shimizu92,Morinaga94}. From a stability
analysis, analytic expressions for the cut-off frequency are found.
For atoms, which have a quadratic Stark shift, the cut-off frequency
is linear in the applied voltage. For high $f$, the trap depth
decreases with $1/f^2$, so the maximum depth is obtained for
frequencies just above the cut off, and its position, $f_{\rm
peak}$, is also expected to scale linear in the applied voltage.

From previous studies~\cite{Junglen04} we know that the regions away
from the center, where there are no known analytic solutions of the
equations of motion, play an important role. Therefore, extensive
Monte Carlo simulations were performed to study the performance of
the trap in detail. In these simulations, the Newtonian equations of
motion are solved for typically $10^4$ particles for a single
parameter setting. All parameters in the simulation are fixed by the
experiment. For the electric field in the simulation, the exact
three-dimensional calculation was used~\cite{Simion}.

Our experimental realization of the trap has titanium electrodes
mounted on a claw-like ceramic mount. The diameter of the electrode
rods is 3\,mm, which is also the distance between opposing
electrodes. The radius of curvature of the half-spherical endings is
1.5\,mm. The electrodes can be charged to $\pm7\,$kV in a time
shorter than a microsecond. The (four) diagonals through the trap
offer free sight with a diameter of 1.9\,mm. The trap is oriented
such as to allow for three optional orthogonal optical paths
parallel and perpendicular to the horizontal plane. Along these
three axes, the free view is 1.4\,mm. Due to the tilted mounting,
gravity is pointing in the $(-0.67,-0.67,-0.33)$ direction of the
cartesian coordinate system coinciding with the axes of the
electrodes.

The electric trap is loaded with laser-cooled $^{85}$Rb atoms. The
atoms are captured from a dispenser in a MOT placed at a horizontal
distance of 2\,cm from the center of the electric trap. As measured
by absorption imaging, $2\times10^7$ atoms are captured here. Before
transferring the atoms to a magnetic trap, they are cooled by
optical molasses~\cite{molasses} to a temperature of
$\approx15\,\mu$K in 7\,ms and optically pumped to the $F=3, m_F=3$
Zeeman state. The magnetic trap is generated by two water-cooled
coils with 48 windings which are rated to hold 75\,A for 3 seconds.
The coils are mounted on a translation stage. By moving the coils,
the magnetic field minimum and the atoms are
translated~\cite{Lewandowski03}. After bringing the atoms in
approximately $0.8\,$s into the center of the electric trap, the
magnetic trap is switched off in $25\,\mu$s and the electric trap is
switched on $100\,\mu$s later. To determine the initial conditions
of the atoms before the start of the electric trap, we kept the
atoms at the position of the MOT and measured the cloud to contain
$6\times10^6$ atoms and to have a temperature of $62\,\mu$K. In a
separate experiment, it was tested that moving the magnetic trap
causes no significant heating or losses.

\begin{figure}[htb]\begin{center}
\epsfig{file=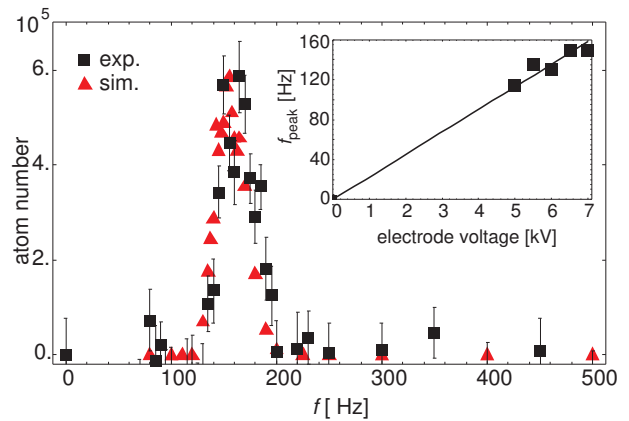,width=0.45 \textwidth}%
\caption{(color online) The frequency dependence of the trapped
number of atoms after 50\,ms for electrodes at $\pm7\,$kV. The
square symbols are data, the triangles are the result of a Monte
Carlo simulation. Noise in the background images can lead to
negative values. The simulation result is scaled to fit the height
of the data. The inset shows the measured dependence of the
frequency of the maximum $f_{\rm peak}$ as a function of the
electrode voltage. The expected linear dependence is indicated by
the line.}
\label{FreqDependence}%
\end{center}\end{figure}
A number of different measurements was performed to characterize the
electric trap: the dependence of the number of trapped atoms on
time, drive frequency and electrode voltage was studied and is
discussed in the following. The number of particles remaining in the
electric trap after 50\,ms was measured by absorption imaging. The
results are depicted in Fig.~\ref{FreqDependence} as a function of
the drive frequency. There is a sharp cut off near 120\,Hz. Just
above the cut off, the maximum number of atoms is trapped. As
expected, the frequency of the maximum, $f_{\rm peak}$, scales
linearly in the applied voltage (see the insert of
Fig.~\ref{FreqDependence}). In the limit of high frequencies, the
micromotion becomes smaller and faster and the net force towards the
trap center vanishes. Therefore, the trap depth, and hence the
number of particles, falls off quickly with $f$. The simulation
describes the position and the shape of the trapping peak very well.
The absolute fraction of trapped atoms is very sensitive to
temperature, size and position of the initial atomic cloud.
Nevertheless, this fraction obtained from the simulation agrees with
the measurement within a factor of two. For trapping times below
50\,ms, the atom number as a function of driving frequency shows
more complex structures (not shown), both in the simulation and in
the experiment, which is attributed to atoms which are released from
the magnetic trap into unstable orbits of the electric trap.

\begin{figure}[htb]\begin{center}
\epsfig{file=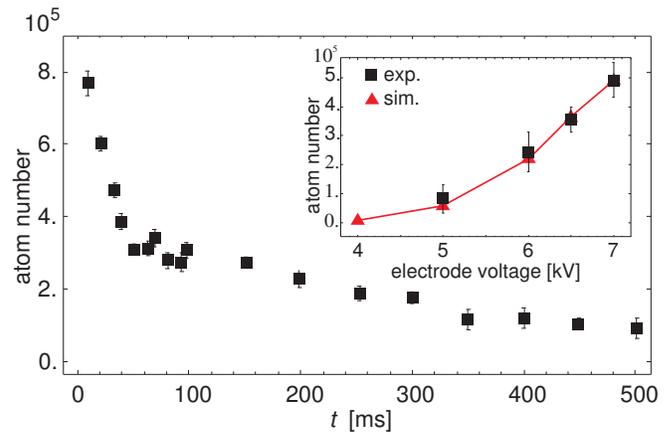,width=0.48\textwidth}%
\caption{(color online) Remaining number of atoms as a function of
time. At $t=0$ the magnetic trap contains $2.7\times10^6$ atoms. The
initial rapid decay is caused by atoms which enter the electric trap
on non-trapped orbits, either because they are too hot or because
their motion has the wrong phase. The decay of the trapped atoms,
visible as the long tail with a $1/e$ time of 0.36\,s, is attributed
to collisions with back-ground gas. The inset shows, for another set
of data with other starting conditions, the dependence of the
maximum number of trapped atoms after 50\,ms electric trapping as a
function of the applied voltage. The line is a guide to the eye.}
\label{lifetime}%
\end{center}\end{figure}

The number of remaining atoms as a function of time is plotted in
Fig.~\ref{lifetime}. The atoms escaping the trap from unstable
orbits causes the loss which is observed for times shorter than
50\,ms. The simulation predicts this rapid initial loss. It also
predicts stable trapping after $50-80\,$ms. The observed signal,
however, falls off with a time constant of $0.36\,$s. The pressure
in the UHV chamber is in the $10^{-10}\,$mbar range during MOT
operation, which is too good to explain this loss. However, during
operation of the electric trap, significant pressure rises are
observed by a remote pressure gauge. Also the lifetime in the
magnetic trap just outside the electrodes is observed to go down
from almost 2\,s to the few 100\,ms regime when high voltage is
applied. From this and other observations we deduce that gas,
probably Rb absorbed on the electrodes, is ejected from the
electrodes once they are charged up, locally increasing the pressure
and limiting the lifetime of trapped atoms. Therefore, we are
confident that the observed decay is due to collisions with hot
background atoms released from the electrodes.

The maximum number of trapped atoms depends not only on the drive
frequency and time but also on the applied voltages. In the inset in
Fig.~\ref{lifetime}, the number of atoms in the maximum is plotted
as a function of the applied voltages. Again, the simulation
describes the results well. Below 4\,kV, trapping is no longer
possible due to the gravitational force exceeding the cycle-averaged
electric trapping force.

\begin{figure}[htb]\begin{center}
\epsfig{file=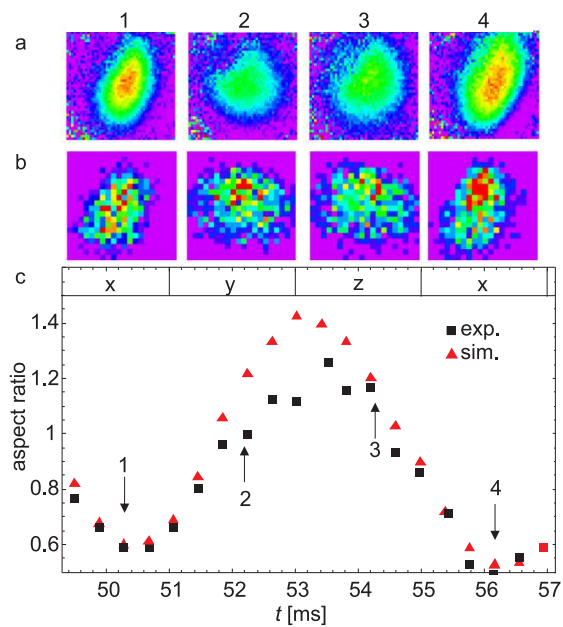,width=0.43 \textwidth}%
\caption{(color online) a) Measured and b) simulated camera pictures
of the trapped cloud at the times indicated $1,2,3,4$ in c). Each
picture in a) is an average over 10 measurements. The $19.5^\circ$
tilt with respect to the vertical corresponds to the tilt angle of
the trap mounting. c) Measured (squares) and simulated (triangles)
aspect ratio of the elliptic clouds. The boxes with letters
$x,y,z,x$ indicate the charged electrodes. Note that the simulation
has no free parameters. }
\label{images}%
\end{center}\end{figure}

Absorption imaging of the trapped cloud allows additional insight
into its dynamics. Images spanning more than a full cycle of the AC
trap are depicted in Fig.~\ref{images}. The atom cloud is
ellipsoidal, but the ellipticity varies periodically in time. This
is caused by the periodic motion of the individual atoms, making the
cloud pulsate. To demonstrate the correspondence between the
simulation and the measurements, we calculate the ellipticity of the
measured and simulated clouds by projection onto their two major
axes. From this the aspect ratio is calculated and plotted in
Fig.~\ref{images}c). The agreement is very good; the remaining
differences are attributed to imaging imperfections. Note the
difference in the minimum ellipticity at 50.5 and 56\,ms, which is
predicted by the simulation and caused by a residual vertical
oscillation of the cloud under influence of gravity with a longer
periodicity.

As a final characterization of the trap, its depth and the
temperature of the trapped cloud should be discussed. From expansion
data, temperatures of the trapped cloud can be obtained. However,
the expansion of the cloud is dominated by the velocities of the
micromotion, which remain constant in the falling frame of reference
when the electric fields are switched off. The ellipticity observed
in Fig.~\ref{images} already indicates that the expansion will be
anisotropic. Indeed, different temperatures are measured depending
on the phase when the trap is switched off. At a driving frequency
of 170\,Hz and release from the electric trap after 21\,ms,
temperatures of $63\,\mu$K and $122\,\mu$K were measured in two
orthogonal directions. One could argue, however, that these numbers
are not good measures for the trap depth. More relevant is the depth
of the effective potential, which in the harmonic limit is obtained
by averaging over the micromotion. Here the simulation is again of
help: it allows to extract the {\it initial} velocity distribution
of those atoms which remain trapped in stable orbits. For a cloud of
an initial size of 0.3\,mm, this distribution corresponds to
$20\,\mu$K. We use this number as a definition of the effective trap
depth.

There are two natural ways to improve the system. The lifetime could
be extended by improving the background pressure during operation of
the electric trap. The MOT could be placed further away from the
electric trap or even in a separate, differentially pumped, chamber.
A second improvement concerns gravity. The simulations show that
without gravity, the trap depth is approximately doubled. This
implies that significantly more particles could be trapped by
compensating for gravity. This can, in principle, be achieved by
applying asymmetric voltages on the electrodes, shifting the saddle
point of the electric field up, causing a net electric force to
outbalance gravity. This is more straightforward when the trap is
rotated such that one of the electrode pairs is oriented vertically.

In principle, simultaneous trapping of atoms and molecules in our
trap is possible, enabling ultracold collision studies once
sufficiently high densities and sufficiently long trapping times are
achieved. Simultaneous trapping should work for molecules like
H$_2$O or D$_2$O which have, firstly, a quadratic Stark
shift~\cite{RiegerPRA06} and, secondly, several thermally populated
states with a similar polarizability over mass ratio as $^{85}$Rb.
Other alkalis and higher-order stability islands could also be
exploited. Another possibility is to overlay a magnetostatic trap
for atoms with a tens of mK deep electrodynamic trap for molecules
with a large Stark shift. Since optimal switching frequencies for
such molecules are in the kHz range~\cite{Junglen04,Veldhoven05},
these fields will only make a very weak potential for the atoms.
Hence, sympathetic cooling of cold molecules by laser-cooled atoms
might be possible. Other applications of our setup are trapping of
atoms or molecules in Rydberg states and the study of atom-atom
interaction at ultralow temperatures by electric
fields~\cite{Marinescu98} or combined electric and magnetic
fields~\cite{Krems05}.

During the final preparation stages of this manuscript, similar
results were reported elsewhere~\cite{Schlunk07}.

We thank S. Nu{\ss}mann for help with the experiment. We acknowledge
financial support by the Deutsche Forschungsgemeinschaft (SPP 1116,
cluster of excellence Munich Centre for Advanced Photonics).


\end{document}